\documentclass[11pt]{article}
\usepackage{jheppub}
\usepackage{amsmath,amsopn,amssymb,amsfonts,amsthm,mathrsfs,bbm,caption,latexsym,stmaryrd,verbatim,color,graphicx}
\usepackage{hyperref}
\usepackage[dvipsnames]{xcolor}

\let\l\left
\let\r\right

\let\mb\mathbb
\let\mc\mathcal

\newcommand{\nn}{\nonumber\\}

\title{Dirac zero modes for Abelian BPS multimonopoles}
\author{J. Lamy-Poirier}
\affiliation{Perimeter Institute for Theoretical Physics, 
Waterloo, Ontario, Canada N2L 2Y5}
\affiliation{Department of Physics and Astronomy,
University of Waterloo, Ontario, Canada N2L 3G1}

\abstract{We develop a method for finding the zero modes of the Dirac operator in the presence of BPS monopoles. We use it to find the zero modes in the case of Abelian BPS monopoles in $\mb R^3$.
}

\begin{document}

\maketitle
\flushbottom

\newpage

\section{Introduction}

In the construction of BPS monopoles one usually consider a related Dirac operator $\mc D^{\dagger}$ and its normalizable zero modes \cite{Nahm:1982jt}. In particular the zero modes play an important role in the Nahm transform \cite{Nahm:1979yw,Nahm:1982jb} used for computing the monopoles, where they provide a solution to the Nahm equations. However, while in many cases the form of the monopoles is known, the zero modes themselves remain hard to compute.

In this paper we aim to develop a method for finding the Dirac zero modes, by considering the case of Abelian (singular) BPS monopoles in $\mb R^3$. These monopoles have a relatively simple field configuration, but their zero modes remain difficult to compute. Only a few results are known, concerning single monopoles \cite{Cheng:2013mla} and certain cases of double monopoles \cite{VanBaal:2002rt}. In this paper we present a general formula for arbitrary (finite) monopole configurations. The formula is explicit (in terms of solutions to a finite set of linear equations) for monopoles of charge $\pm1$ at generic positions, while solutions for the other cases can be recovered by a limiting process. The formula takes the form of a sum of residues at a certain set of poles, where the location of the poles corresponds to the zeros of algebraic functions.

Our method is based on the fact that a large class of (generally non-normalizable) solutions to the equation $\mc D^{\dagger}\Psi=0$ can be built from flat sections of a Lax pair of connections. We conjecture that the zero modes can be written as superpositions of such solutions, or equivalently that the zero modes are of the form $\mc D \chi$, where $\mc D$ is the adjoint of $\mc D^{\dagger}$. In the Abelian case, the flat sections are easily found, so the conjecture reduces the problem to finding which superpositions give the correct zero modes, i.e. which ones are normalizable. It is worth noting that while the conjecture itself appears to be new, some similar ideas can be found in the literature. In \cite{Nahm:1982jt} the zero modes for smooth monopoles are related to flat sections of the Lax pair in a different way, in terms of a differential equation involving normalizable flat sections. Also the dipole solution in \cite{VanBaal:2002rt} is found in the form $\mc D \chi$.

Our main motivation for computing Dirac zero modes is to obtain the BPS spectrum of two-dimensional gauged linear sigma models (GLSM), and by extension that of many non-linear sigma models (NLSM). The BPS spectrum of GLSM is encoded in a cylindrical geometry, through a topological-anti-topological metric \cite{Cecotti:1991me} and a supersymmetric index \cite{Cecotti:1992qh} (see also \cite{Cecotti:1992rm}). The BPS spectrum is independent of the radius of the cylinder, and for a small radius we obtain an effective one-dimensional theory, corresponding to the quantum mechanics of vector multiplets in the presence of a background of periodic monopoles. In particular the supersymmetric index of \cite{Cecotti:1992qh} can be computed in terms of the ground states of the resulting theory, which correspond precisely to the Dirac zero modes for periodic monopoles. While in this paper we only consider non-periodic monopoles, we hope that our results can be generalized to the periodic case.

The paper is organized as follow. In section \ref{review} we briefly review some background material on Abelian BPS monopoles and the associated Dirac operator. In section \ref{Lax} we review the basic properties of the Lax pair for the Bogomolnyi equations, and show how it gives an ansatz for the Dirac zero modes. We continue in section \ref{Example} with the simplest example, a single monopole of unit charge, and use the ansatz to find the correct zero mode. The result takes the form of a residue at a special value of a spectral parameter. In the following section we show that the formula for a generic monopole configuration with positively charged monopoles can also be expressed as a residue formula. We find the exact formula for monopoles for monopoles of unit charge at generic positions, and the generalization to other cases is straightforward. We leave the details of the computation for appendix \ref{Proof}, and we generalize the results for negative charges in appendix \ref{negative}. In section \ref{Nahm} we review some of the implications of our results for the Nahm transform, and give an expression for the solutions of the relevant Nahm equations in terms of an integral over an algebraic variety. We conclude in section \ref{Conclusion} with some open questions concerning the results of this paper and their possible generalizations.

\section{BPS monopoles and Dirac zero modes}\label{review}

A BPS monopole configuration on $\mb R^3$  with gauge group $G$ consists of a vector field ${\bf A}({\bf x})$ and a Higgs field $\Phi({\bf x})$, satisfying the Bogonolnyi equation
\begin{align}
{\bf D}\Phi=*{\bf F},
\end{align}
where ${\bf D}={\bf d}-i{\bf A}$ is the connection defined by ${\bf A}$, and ${\bf F}$ is the curvature of ${\bf A}$. The configuration must also be sufficiently regular at infinity. In this paper we restrict the gauge group to $G=U(1)$. In this case a monopole configuration corresponds to a set $S$ of monopoles, described by their charge $q\in \mb Z^*$ and their location ${\bf a}\in\mb R^3$. 
We can choose a gauge such that the fields are given by \cite{Kronheimer:1985,Cheng:2013mla}
\begin{align}
\Phi({\bf x})=\sum_{m\in S}\frac{q_m}{2r_m},\qquad
{\bf A}({\bf x})=i\sum_{m\in S}\frac{z_m{\bf d}\bar z-\bar z_m {\bf d}z}{4r_m(r_m-x_m)}.
\end{align}
Here we work in the coordinates $x=x^3$, $z=x^1+ix^2$, and we write ${\bf x}_m={\bf x}-{\bf a}_m$, $r_m=||{\bf x}_m||$.

In this paper we are interested in the Dirac operators
\begin{align}
\mc D^\dagger={\boldsymbol \sigma}\cdot{\bf D}+\Phi-t,\qquad  \mc D={\boldsymbol \sigma}\cdot{\bf D}-\Phi+t,
\end{align}
where $t\in \mb R$. Specifically we are looking for zero modes of $\mc D^\dagger$ in the Hilbert space $\mc H$ of spinors $\Psi$ satisfying $\mc D^{\dagger}\Psi=0$, which are non-singular on $\mb R^3\backslash\{{\bf a}_m\}$ (up to gauge-dependent phase singularities) and square-integrable:
\begin{align}
\int_{\mb R^3} d^3x \bar\Psi\Psi<\infty.
\end{align}
It is expected that the number of zero modes is $N_+=\sum_{m\in S_+}q_m$ for $t\in\mb R_+$, and $N_-=\sum_{m\in S_-}|q_m|$ for $t\in\mb R_-$, where $S_\pm=\{i|q_i\in\mb Z_\pm\}$. Indeed, Abelian monopoles can be obtained from a large Higgs field limit of smooth $SU(2)$ monopoles, for which a similar result holds by an index theorem \cite{Hitchin:1988hk,Nakajima:1990}. The above claim is also predicted by string theoretical constructions of the Nahm transform \cite{Diaconescu:1996rk,Cherkis:1997aa}.

\section{Twistor space, Lax pair and the Dirac equation}
\label{Lax}

In the following we will consider the space $T$ of oriented straight lines in $\mb R^3$, which is identical to the holomorphic tangent space $T\mb P^1$ of the (complex) projective line $\mb P^1$ \cite{Hitchin:1982gh}. A line in $T$ can be described (non-uniquely) by a base point $\bf x$ and a direction $\bf v$. It is useful to view the direction as a point in $\mb P^1$, represented by the coordinate 
\begin{align}\label{Connection}
\zeta({\bf v})=\frac{||{\bf v}||+v^3}{v^{\bar z}}
\end{align}
For any $\zeta$, there is a convenient choice of coordinates for $\bf x$:
\begin{align}
u(\zeta;{\bf x})=\frac{z}{2\zeta}+\frac{\bar z\zeta}{2},\qquad
v(\zeta;{\bf x})=-\frac{z}{2\zeta}+\frac{\bar z\zeta}{2}+x,\qquad
y(\zeta;{\bf x})=\frac{z}{2\zeta}-\frac{\bar z\zeta}{2}+x.
\end{align}
The space of lines passing through a point $\bf x$ forms a real section of $T\mb P^1$, noted $P_{\bf x}$. Given two distinct points $\bf x$, $\bf y$, we denote by $C_{\bf x}\cdot C_{\bf y}$ the line passing through both points, passing through $\bf x$ first.

In the study of singular monopoles, it is useful to consider spaces of half-lines in $\mb R^3$ \cite{Kronheimer:1985}. We define $T^+$ as the space of oriented half-lines pointing away from their endpoint, and similarly we define $T^-$ from half-lines pointing towards their endpoint\footnote{Alternatively, $T^+$ is built from linear maps $(0,\infty)\to\mb R^3$, while $T^-$ comes from maps $(-\infty,0)\to\mb R^3$}. Given a point $\bf x$, we write the spaces of lines ending at $\bf x$ as $C^\pm_{\bf x}$.

Given $\zeta\in\mb C$, one has special pair of connections \cite{Nahm:1982jt},
\begin{align}\label{Connection}
\nabla_\zeta&=\zeta D_z+\zeta^{-1}D_{\bar z}-\Phi+t\equiv D_{u(\zeta)}-\Phi+t,\nn
\tilde\nabla_\zeta&=\zeta D_z-\zeta^{-1}D_{\bar z}+D_3\equiv 2D_{v(\zeta)},
\end{align}
forming a Lax pair for the Bogonolnyi equation. Indeed, the flatness condition for ($\nabla_\zeta$, $\tilde\nabla_\zeta$) is equivalent to the Bogonolnyi equation. Here we are interested in the flat sections $\chi({\bf x},t;\zeta)$ of the Lax pair, satisfying
\begin{align}\label{CR}
\nabla_\zeta\chi({\bf x},t;\zeta)=0,\qquad \tilde\nabla_\zeta\chi({\bf x},t;\zeta)=0.
\end{align}

The connections (\ref{Connection}) can be related to the Dirac operators $\mc D$ and $\mc D^\dagger$ as follow. We make the ansatz 
\begin{align}
\tilde\Psi({\bf x},t;\zeta)=\chi({\bf x},t;\zeta) \l(
\begin{matrix}1\\\zeta\end{matrix}
\r).
\end{align}
Then the equation $\mc D\tilde\Psi=0$ reduces to\footnote{Here and in the following we relax the assumption that the spinors lie in the Hilbert space $\mc H$, and allow for more singular functions. Hence ``solving'' the equation $\mc D\tilde\Psi=0$ makes sense here even though none of the solutions lie in $\mc H$. The assumption will be restored later, as a condition for the zero modes of $\mc D^\dagger$ to lie in $\mc H$.}
\begin{align}
0=\mc D\tilde\Psi&=\l(
\begin{matrix}
D_3-\Phi+t&
2D_z\\
2D_{\bar z}&
-D_3-\Phi+t
\end{matrix}
\r)
\chi({\bf x},t;\zeta) \l(
\begin{matrix}1\\\zeta\end{matrix}
\r)\nn
&=
\l(
\begin{matrix}
2\zeta D_z+D_3-\Phi+t&
\\
\zeta(2\zeta^{-1}D_{\bar z}
-D_3-\Phi+t)
\end{matrix}
\r)\chi({\bf x},t;\zeta),
\end{align}
which is equivalent to  eq. (\ref{CR}).
Thus given a set of solution to eq. (\ref{CR}), one can obtain 
solutions to $\mc D\tilde\Psi=0$ by integrating over $\zeta$ against arbitrary functions $F(\zeta)$. (This corresponds to an integral over $C_{\bf x}$.) However, here we are interested in the zero modes of $\mc D^\dagger$. We can actually build a large class of solutions to $\mc D^\dagger\Psi=0$ from those of eq. (\ref{CR}), of the form
\begin{align}
\Psi({\bf x},t;\zeta)=\mc D\l(
\begin{matrix}\chi({\bf x},t;\zeta) \\0\end{matrix}
\r).
\end{align}
Since $\mc D^\dagger \mc D=({\bf D}^2-(\Phi-t)^2)\otimes \mb I_2$ is diagonal, it follows that $\mc D^\dagger\Psi({\bf x},t;\zeta)=0$. We now claim that all the zero modes in $\mc H$ can be obtained as combinations of such solutions. In the rest of this paper we prove this statement for $U(1)$ BPS monopoles in $\mb R^3$ by finding an explicit expression for $\chi({\bf x},t;\zeta)$.

The approach of this section is in many points similar to that of \cite{Nahm:1982jt}, although in that paper only smooth monopoles are considered. However in that paper the zero modes of $\mc D^\dagger$ are obtained in a different way: given a square integrable 
function $\eta$ satisfying $\mc D\eta=0$, a zero mode $\Psi$ is built from the equation ${\boldsymbol\nabla}\cdot(\Psi^\dagger{\boldsymbol\sigma}\eta)=0$. It would be interesting to relate the two approaches.

\section{Simple example: single monopole of unit charge}\label{Example}

We now proceed to a simple example of the method outlined above and write a known zero mode in terms of flat sections of the Lax connection. We consider a single monopole of unit charge centered at the origin. Assuming $t>0$, the unique zero mode \cite{Cheng:2013mla} of $\mc D^\dagger$ in $\mc H$ is 
\begin{align}\label{singlemonopole}
\Psi({\bf x},t)=\frac{e^{-rt}}r
\l(\begin{matrix}
\sqrt{r-x}\\
-e^{i\phi}\sqrt{r+x}
\end{matrix}\r)
=\mc D\frac{e^{-rt}+ce^{-xt}}{t\sqrt{r-x}}
\l(\begin{matrix}
1\\
0
\end{matrix}\r),
\end{align}
where $c$ is an arbitrary constant which we set to zero. The flat sections are given by
\begin{align}
\chi({\bf x},t;\zeta;F)=\frac{(r-x)^{1/2}}{\bar z \zeta+r-x}e^{tu(\zeta)}
F(\zeta,y(\zeta)),
\end{align}
where $F$ is an arbitrary function. In this simple case we can recover the zero mode (\ref{singlemonopole}) by guesswork: letting $F(\zeta,y)=\zeta^{-1}$ and taking the residue at $\zeta_0=\zeta(-{\bf x})=-(r-x)/\bar z$, we find
\begin{align}
\frac{(r-x)^{1/2}}{2\pi i\bar z}\oint_{\zeta_0} \frac{d\zeta }{\zeta(\zeta-\zeta_0)}e^{tu(\zeta)}
=\frac{i(r-x)^{1/2}}{\bar z \zeta_0}e^{-rt}=-\frac{e^{-rt}}{\sqrt{r-x}},
\end{align}
i.e.
\begin{align}
\Psi({\bf x},t)=-\frac{1}{2\pi i t}\mc D \oint_{\zeta_0}\chi({\bf x},t;\zeta;\zeta^{-1})
\l(\begin{matrix}
1\\
0
\end{matrix}\r).
\end{align}

\section{The general formula for Abelian BPS monopoles on $\mb R^3$}\label{Formula}

We now show that the residue formula is quite general, and in fact can be generalized to any set of $U(1)$ BPS monopoles. Here we find an exact formula for monopoles of unit charge at generic positions, and as outlined at the end of this section, special cases and generic positives charges can be obtained in a similar way. In appendix \ref{negative} we explore the case where negative charges are also included. In this section, we assume $t>0$.

We consider a set of $N$ monopoles of unit charges $q_m=1$, at positions ${\bf a}_m$ ($m=1,\cdots,N$). We assume that the positions are generic, which requires the following:
\begin{itemize}
\item No pair of monopoles are separated only by a translation in the $x^3$ direction, i.e. ${\bf a}_m-{\bf a}_n$ is not proportional to ${\bf \hat e}^3$ for any $m\neq n$ (this prevents issues with the choice of gauge).
\item No triplet of positions lie on the same line.
\end{itemize}
In particular, all the positions are different. We now make the following definitions:
\begin{align}
{\bf x}_{m}&={\bf x}-{\bf a}_m,\qquad \text{(similarly for $x_m$, $z_m$, $\bar z_m$, $r_m$, $u_m$, $v_m$, $y_m$)},\nn
{\bf a}_{mn}&={\bf a}_m-{\bf a}_n,\qquad a_{mn}=||{\bf a}_{mn}||
\end{align}
Flat sections of the Lax connection take the form
\begin{align}
\chi({\bf x},t;\zeta;F)&=\l(\prod_m\chi_0({\bf x}_m)\r)e^{tu(\zeta)}
F(\zeta,y(\zeta)),\nn
\chi_0({\bf x})&=\frac{(r-x)^{1/2}}{\bar z(\zeta-\zeta(-{\bf x}))}=\l(\frac{-\zeta(-{\bf x})}{\bar z}\r)^{1/2}(\zeta-\zeta(-{\bf x}))^{-1}.
\end{align}
To simplify the twistor notation, we define
\begin{align}
\zeta_m=\zeta(-{\bf x}_m)=-\frac{r_m-x_m}{\bar z_m}, \qquad
\zeta'_m=\zeta({\bf x}_m)=\frac{r_m+x_m}{\bar z_m}, \qquad
\zeta_{mn}=\zeta({\bf a}_{nm}).
\end{align}
Note that the coordinates $\zeta_m({\bf x})$ and $\zeta_m'({\bf x})$ correspond to the zeros of $y(\zeta;{\bf x}_m)$:
\begin{align}
y(\zeta;{\bf x})&=-\frac{\bar z}{2\zeta}(\zeta-\zeta_m)(\zeta-\zeta'_n).
\end{align}
Inspired by the results of the previous section, we seek a residue formulation for the zero modes. 
We pick a residue at each of the $\zeta_m$, corresponding to the direction of the monopoles\footnote{
Here one could try using more complicated residues, but these are the only choices allowing square-integrability at infinity in ${\bf x}$. For example if we divide by a power of $y(\zeta;{\bf x}_m)$ to obtain a higher order pole at $\zeta_m$, we also create a pole at $\zeta_m'$. That additional pole must then also lie inside the integration contour, as the integration contour cannot depend explicitly on ${\bf x}$ (and a contour depending solely on $y(\zeta)$ cannot distinguish the two poles). However $e^{tu(\zeta_m')}=e^{t r_m}$ grows exponentially at infinity, so the result is not square-integrable. One could also try including in $F(\zeta, y)$ a pole at some other location and integrating around it, however this can be ruled out by a similar argument.
}, and define

\begin{align}\label{Residue}
X_n[F]&\equiv\frac1{2\pi i}\oint_{\zeta_{n}}d\zeta\l(\prod_m\chi_0({\bf x}_m)\r)e^{tu(\zeta)}\zeta^{-1}F(\zeta)\nn
&=\l(\prod_m\frac{-\zeta_m}{\bar z_m}\r)^{1/2}
\frac{e^{-r_nt}\zeta_{n}^{-1}F(\zeta_{n})}{\prod_{m\neq n}(\zeta_{n}-\zeta_{m})},\nn
X[\{F_n\}]&=\sum_nX_n[F_n]=\l(\prod_m\frac{-\zeta_m}{\bar z_m}\r)^{1/2}\sum_n
\frac{e^{-r_nt}\zeta_{n}^{-1}F_n(\zeta_{n})}{\prod_{m\neq n}(\zeta_{n}-\zeta_{m})}.
\end{align}
Note that we dropped the second argument of $F$, since $y(\zeta_n;{\bf x})$ is a constant in ${\bf x}$. We now look for zero modes in $\mc H$ of the form
\begin{align}
\Psi[\{F_n\}]&=\mc D 
\l(\begin{matrix}
X[\{F_n\}]\\
0
\end{matrix}\r).
\end{align}
This can be achieved by sets of functions $\{F_n\}$ of the form
\begin{align}\label{solF}
F_p(\zeta)&=\prod_{m\neq p}(\zeta-\zeta_{pm})A_p+\sum_{n\neq p}e^{-a_{pn}t}F_n(\zeta_{pn})\prod_{m\neq n,p}\frac{\zeta-\zeta_{pm}}{\zeta_{pn}-\zeta_{pm}},
\end{align}
for any set of constants $A_p$. The values $F_n(c_{pn})$ can be fixed by the consistency conditions
\begin{align}\label{consistency}
F_p(\zeta_{qp})&=\prod_{m\neq p}(\zeta_{qp}-\zeta_{pm})A_p+\sum_{n\neq p}e^{-a_{pn}t}F_n(\zeta_{pn})\prod_{m\neq n,p}\frac{\zeta_{qp}-\zeta_{pm}}{\zeta_{pn}-\zeta_{pm}}.
\end{align}
This leaves $N$ free parameters $A_p$, or $N$ zero modes\footnote{Here we assumed without proof that the set of equations (\ref{consistency}) is linearly independent. While this can be justified by the expected number of zero modes, it would be interesting to give a concrete proof of the statement. 
}. In appendix \ref{Proof} we derive eq. (\ref{solF}) and prove that it generates the correct zero modes. 

The solutions in the non-generic cases can be obtained by taking a careful limit of the above equations (or by adapting the analysis of this section or appendix \ref{Proof}). In particular, monopoles of higher charges can be obtained from monopoles at coincident position. Note that in that case the integrands in the residue formula have multiple poles.

\section{Relation to Nahm's formulation}\label{Nahm}

The zero modes found in this paper allow to compute solutions to the Nahm equations and the Nahm zero modes via the Nahm transform. In this section we briefly review the procedure, following \cite{Nahm:1982jb,Nahm:1982jt}. 
In this section we relate the results of this paper to Nahm's approach to monopoles. 

Given a set $\{\Psi_i\}$ of zero modes of $\mc D^\dagger$, one forms the matrices
\begin{align}
{\bf T}(t)_{ij}=-i\int_{\mb R^3} d^3x \Psi_i({\bf x},t)^\dagger {\bf x}\Psi_j({\bf x},t),\label{Nahm transform}\nn
T^0(t)_{ij}=-i\int_{\mb R^3}  d^3x \Psi_i({\bf x},t)^\dagger \partial_t\Psi_j({\bf x},t).
\end{align}
We can always set $T^0=0$ by a suitable gauge transformation. Assuming this, the other three matrices satisfy the Nahm equations
\begin{align}
\partial_t{\bf T}={\bf T}\times {\bf T}.
\end{align}
Associated to these matrices is a pair of operators
\begin{align}
D^\dagger=-i\partial_t+{\boldsymbol\sigma}\cdot({\bf T}+{\bf x}),\qquad 
D=i\partial_t+{\boldsymbol\sigma}\cdot({\bf T}+{\bf x}),
\end{align}
playing a role similar to $\mc D^\dagger$, $\mc D$. The zero modes $v_i({\bf x},t)$ of $D^\dagger$ can be related to those of $\mc D^\dagger$ by the formulas
\begin{align}
\Psi_i({\bf x},t)=-\frac{1}{2\sqrt{2\pi}}\epsilon\mc D^\dagger\mc D v_i({\bf x},t)^\dagger,\qquad
v_i({\bf x},t)=-\frac{\sqrt{2\pi}}{2}\epsilon (D^\dagger D)_{ij}\Psi_j({\bf x},t),
\end{align}
where $\epsilon$ is the two-dimensional antisymmetric tensor.

\subsection{An integral formulation}

The results of this paper allow to write solutions of the Nahm equation via eq. (\ref{Nahm transform}), although the integral is in general difficult to compute. We can however use the residue formula to write the solutions as sets of integrals over algebraic varieties. In  the following we explore the formulation of these integrals, but some additional work is needed in order to understand their meaning and relevance. 

Schematically the integrals take the form\footnote{Here the ``Dirac delta functions'' are just schematic, as we do not define the integration contour properly.}
\begin{align}
{\bf T}(t)_{ij}&=-i\sum_{m,n}\int_{\mb R^3} d^3x\oint_{\zeta_m^*}d\tilde\zeta\oint_{\zeta_n}d\zeta \bar\chi_i({\bf x},t,\tilde\zeta)^\dagger {\bf x}\chi_j({\bf x},t,\zeta)\nn
&=-i\sum_{m,n}\int d^3xd\tilde\zeta d\zeta \bar\chi_i({\bf x},t,\tilde\zeta)^\dagger {\bf x}\chi_j({\bf x},t,\zeta)\delta(\tilde\zeta-\zeta_m^*)\delta(\zeta-\zeta_n)\nn
&=-i\sum_{m,n}\int d^3xd\tilde\zeta d\zeta \bar\chi_i({\bf x},t,\tilde\zeta)^\dagger {\bf x}\chi_j({\bf x},t,\zeta)\bar y_m(x,\tilde\zeta)y_n(x,\zeta)\delta(\bar y_m(x,\tilde\zeta))\delta(y_n(x,\zeta)),
\end{align}
for some spinors $\chi_i({\bf x},t,\zeta)$ (which include the action of the Dirac operator $\mc D$). Therefore we have an integral over the algebraic varieties defined by $\bar y_m(x,\tilde\zeta)=y_n(x,\zeta)=0$ in the space spanned by ($\bf x$, $\tilde\zeta$, $\zeta$), and the integral is nontrivial only on the components defined by 
$\tilde\zeta=\zeta_m^*$, $\zeta=\zeta_n$. For $n=m$ we can identify $\tilde\zeta=\zeta^*$, and the algebraic variety is the tautological line bundle over $C_{{\bf a}_m}$, and the contributing ``component'' is a half-line bundle over $C^-_{{\bf a}_m}$. The story is similar for $n\neq m$, but in that case we need to consider a complexification of the space of lines. While this formulation does not simplify the computation of the integral, it gives an algebraic interpretation of the integration contour.

\section{Conclusion and open questions}\label{Conclusion}

In this paper we found the Dirac zero modes for Abelian BPS monopoles on $\mb R^3$. Our method relies on the following ideas and results: 
\begin{itemize}
\item A complete set of flat sections for the Lax pair of connections ($\nabla_\zeta$, $\tilde\nabla_\zeta$), providing a large set of solutions to the Dirac equation $\mc D^{\dagger}\Psi=0$.
\item The assumption that the zero modes belong to the above set of solutions.
\item An integration contour, which gives a set of residue at specific points in $\mb P^1$ corresponding to the direction of each of the monopoles, and described by (one of the roots of) the algebraic equations $y(\zeta,{\bf x}_m)=0$. The function $y(\zeta,{\bf x}_m)$ describe the algebraic space $C_{{\bf a}_m}$, the space of lines passing through the monopole.
\item An explicit evaluation of the singularities for the solutions of the Dirac equations obtained from the contour integrals, and a resulting simple set of algebraic conditions for the integrand of the contour integral.
\end{itemize}
As we look for generalizations of the results of this paper, some of the above ideas are bound to fail. For instance the algebraicvarieties $C_{{\bf a}_m}$ considered in this paper are specific to singular monopoles, hence a different integration contour is needed for non-Abelian monopoles. A natural guess is the spectral curve \cite{Hitchin:1982gh,Hitchin:1983ay}. For periodic monopoles the flat section for the Lax pair have an essential singularity near the monopoles, and its cancellation is highly nontrivial. We hope that these difficulties can be overcome.

\acknowledgments{This work was supported by the Perimeter Institute for Theoretical Physics and the Natural Sciences and Engineering Research Council of Canada (NSERC).  Research at Perimeter Institute is supported by the Government of Canada through Industry Canada and by the Province of Ontario through the Ministry of Economic Development and Innovation. The author is thankful to D. Gaiotto and B. Charbonneau for useful discussions and remarks. }

\begin{appendix}

\section{Proof of the general formula}\label{Proof}

In this appendix we prove that the zero modes described in section \ref{Formula} do belong to the Hilbert space $\mc H$. We first derive equation (\ref{solF}) by imposing square integrability near the monopoles, then we show that the singularities at $\zeta_{p}=\zeta_{q}$ all cancel. The ``Dirac string'' type singularities also cancel, however the proof is tedious but straightforward and we do not write it here. Also, it is clear that the solution is regular at infinity if and only if $t>0$.

\subsection{Near the monopoles}

In the first step we impose square integrability near the monopoles. This requires $\Psi[\{F_n\}]$ to be less divergent that $r_n^{-3/2}$. In components, we have
\begin{align}\label{Components}
\Psi_1[\{F_n\}]&=\l(\prod_m\frac{-\zeta_m}{\bar z_m}\r)^{1/2}
\sum_n\frac{e^{-r_nt}\zeta_{n}^{-1}F(\zeta_{n})}{\prod_{m\neq n}(\zeta_{n}-\zeta_{m})}\nn
&\quad\times\l(\sum_{m\neq n}\zeta_{m}^{-1}\partial_3\zeta_{m}+\frac{F_n'(\zeta_{n})}{F_n(\zeta_{n})}\partial_3\zeta_{n}
-\sum_{m\neq n}\frac{\partial_3\zeta_{n}-\partial_3\zeta_{m}}{\zeta_{n}-\zeta_{m}}+t\l(1-\frac{x_n}{r_n}\r)\r),\nn
\Psi_2[\{F_n\}]&=\l(\prod_m\frac{-\zeta_m}{\bar z_m}\r)^{1/2}
\sum_n\frac{e^{-r_nt}\zeta_{n}^{-1}F(\zeta_{n})}{\prod_{m\neq n}(\zeta_{n}-\zeta_{m})}\nn
&\quad\times\l(
\sum_{m\neq n}\zeta_{m}^{-1}\bar\partial\zeta_{m}+\frac{F_n'(\zeta_{n})}{F_n(\zeta_{n})}\bar\partial\zeta_{n}
-\sum_{m\neq n}\frac{\bar\partial\zeta_{n}-\bar\partial\zeta_{m}}{\zeta_{n}-\zeta_{m}}-t\frac{z_n}{2r_n}
\r).
\end{align}
Near a monopole, say $r_p\to0$, only the terms proportional to $\partial_3\zeta_{p}$ or $\bar\partial\zeta_{p}$ contribute to the leading singularity, so we have
\begin{align}
\Psi_1[\{F_n\}]&\sim
\l(\prod_{m\neq p}\frac{-\zeta_{pm}}{(a_{pm})_{\bar z}}\r)^{1/2}\l(\frac{-\zeta_p}{\bar z_p}\r)^{1/2}\zeta_p^{-1}\partial_3\zeta_{p}
\nn
&\times\frac{\partial}{\partial\zeta_p}\l(\frac{F_p(\zeta_{p})}{\prod_{m\neq p}(\zeta_{p}-\zeta_{pm})}
-\sum_{n\neq p}\frac{e^{-a_{pn}t}F_p(\zeta_{pn})}{(\zeta_{p}-\zeta_{pn})\prod_{m\neq n,p}(\zeta_{pn}-\zeta_{pm})}
\r),\nn
\Psi_2[\{F_n\}]&\sim
\l(\prod_{m\neq p}\frac{-\zeta_{pm}}{(a_{pm})_{\bar z}}\r)^{1/2}\l(\frac{-\zeta_p}{\bar z_p}\r)^{1/2}\zeta_p^{-1}\bar\partial\zeta_{p}
\nn
&\times\frac{\partial}{\partial\zeta_p}\l(\frac{F_p(\zeta_{p})}{\prod_{m\neq p}(\zeta_{p}-\zeta_{pm})}
-\sum_{n\neq p}\frac{e^{-a_{pn}t}F_p(\zeta_{pn})}{(\zeta_{p}-\zeta_{pn})\prod_{m\neq n,p}(\zeta_{pn}-\zeta_{pm})}
\r).
\end{align}
Since the leading singularity is of order $r_p^{-3/2}$, it must cancel, leading to the differential equations
\begin{align}
\frac{\partial}{\partial\zeta_p}\l(\frac{F_p(\zeta_{p})}{\prod_{m\neq p}(\zeta_{p}-\zeta_{pm})}
-\sum_{n\neq p}\frac{e^{-a_{pn}t}F_p(\zeta_{pn})}{(\zeta_{p}-\zeta_{pn})\prod_{m\neq n,p}(\zeta_{pn}-\zeta_{pm})}
\r),
\end{align}
solved by (\ref{solF}).

\subsection{Cancellation of singularities: $\zeta_{p}=\zeta_{q}$}

We first consider the case where $\zeta_{p}=\zeta_{q}$ for some $p$, $q$. This happens on a subset of the line connecting ${\bf a}_p$ and ${\bf a}_n$, on the two half lines between ${\bf a}_p$ and ${\bf a}_q$, and $\infty$ in the opposite direction (i.e. outside the interval between the two points). Consider the case of the (open) half line between ${\bf a}_p$ and $\infty$, where $\zeta_{p}=\zeta_{q}=\zeta_{pq}$, the other case being similar. We want to show that $\Psi[\{F_n\}]$ is regular there for any solution of eq. (\ref{solF}). In fact $X[\{F_n\}]$ is also regular there. To show this, we write
\begin{align}
X[\{F_n\}]&=\l(\prod_m\frac{-\zeta_m}{\bar z_m}\r)^{1/2}\l(
\frac{1}{\zeta_{p}-\zeta_{q}}\l[\frac{e^{-r_pt}F_p(\zeta_{p})}{\prod_{m\neq p,q}(\zeta_{p}-\zeta_{m})}
-\frac{e^{-r_qt}F_q(\zeta_{q})}{\prod_{m\neq p,q}(\zeta_{q}-\zeta_{m})}
\r]+\cdots\r),
\end{align}
where $\cdots$ is regular on the half line, while the term inside the square brackets evaluates to
\begin{align}
\frac{e^{-r_pt}F_p(\zeta_{pq})-e^{-r_qt}F_q(\zeta_{pq})}{\prod_{m\neq p,q}(\zeta_{pq}-\zeta_{m})}
=\frac{e^{-r_pt}(F_p(\zeta_{pq})-e^{-a_{pq}t}F_q(\zeta_{pq}))}{\prod_{m\neq p,q}(\zeta_{pq}-\zeta_{m})}.
\end{align}
Meanwhile, evaluating eq. (\ref{solF}) at $\zeta=c_{pq}$ gives $F_p(c_{pq})=e^{-a_{pq}t}F_q(c_{pq})$, so $X[\{F_n\}]$ is regular on the half line. Since the Dirac operator $\mc D$ cannot add a singularity there, it implies $\Psi[\{F_n\}]$ is also regular there.

\section{Negative charges}\label{negative}

In this appendix we generalize the residue formula for a configuration including both monopoles of charge $+1$ and $-1$ (and by appropriate limits any combination any set of integer charges). We consider a configuration containing $N_+$ monopoles of charge $+1$ at positions ${\bf a}_m$, $m\in S_+$, and $N_-$ monopoles of charge $-1$ at ${\bf a}_{\hat m}$, $\hat m\in S_-$. We write the contribution of the negatively charged monopoles to the flat section of the Lax connection in terms of
\begin{align}
\chi_0'({\bf x};\zeta)&=\l(-2\chi_0y(\zeta)\r)^{-1}=\l(-\zeta(-{\bf x})\bar z\r)^{-1/2}(\zeta-\zeta({\bf x}))^{-1}
\end{align}
We define the residues
\begin{align}
X_n[F]&\equiv\oint_{\zeta_{n}}d\zeta\l(\prod_m\chi_0({\bf x}_m)\r)\l(\prod_{\hat m}\chi_0'({\bf x}_{\hat m})\r)e^{tu(\zeta)}\zeta^{-1}F(\zeta),\nn
X_{\hat n}[F]&\equiv\oint_{\zeta_{\hat n}}d\zeta\l(\prod_m\chi_0({\bf x}_m)\r)\l(\prod_{\hat m}\chi_0'({\bf x}_{\hat m})\r)e^{tu(\zeta)}\zeta^{-1}F(\zeta),\nn
X[\{F_n\}]&=\sum_nX_n[F_n],\qquad \hat X[\{F_{\hat n}\}]=\sum_{\hat n}X_{\hat n}[F_{\hat n}],
\end{align}
Note that $X$ and $\hat X$ cannot be mixed, since regularity at infinity requires $t>0$ for $X$ and $t<0$ for $\hat X$. Proceeding as before, we impose square-integrability for $X$ near the monopoles, and find the set of equations
\begin{align}
0&=\frac{\partial}{\partial\zeta_p}\l[\frac{1}{\prod_{\hat m}(\zeta_{p}-\zeta_{\hat mp})}\l(\frac{F_p(\zeta_{p})}{\prod_{m\neq p}(\zeta_{p}-\zeta_{pm})}
-\sum_{n\neq p}\frac{e^{-a_{pn}t}F_p(\zeta_{pn})}{(\zeta_{p}-\zeta_{pn})\prod_{m\neq n,p}(\zeta_{pn}-\zeta_{pm})}
\r)\r],\nn
0&=F_p(\zeta_{\hat m p}).
\end{align}
The first equation is solved by 
\begin{align}
F_p(\zeta)&=\l(\prod_{\hat m}(\zeta_{p}-\zeta_{\hat mp})\r)\l(\prod_{m\neq p}(\zeta-\zeta_{pm})A_p+\sum_{n\neq p}e^{-a_{pn}t}F_n(\zeta_{pn})\prod_{m\neq n,p}\frac{\zeta-\zeta_{pm}}{\zeta_{pn}-\zeta_{pm}}\r),
\end{align}
which trivially satisfies the second. A similar analysis can be performed for $\hat X$, and as before we can show that the formula gives valid zero modes. Also the number of zero modes is $N_+$ for $t>0$, and $N_-$ for $t<0$, as expected.

\end{appendix}

\newpage
\bibliographystyle{JHEP}
\bibliography{irr}

\providecommand{\href}[2]{#2}\begingroup\raggedright\begin{thebibliography}{10}

\bibitem{Nahm:1982jt}
W.~Nahm, {\it {The Algebraic Geometry of Multi-monopoles}},  in {\em {11th
  International Colloquium on Group Theoretical Methods in Physics (GROUP 11)
  Istanbul, Turkey, August 23-28, 1982}}, 1982.

\bibitem{Nahm:1979yw}
W.~Nahm, {\it {A Simple Formalism for the BPS Monopole}},  {\em Phys. Lett.}
  {\bf B90} (1980) 413.

\bibitem{Nahm:1982jb}
W.~Nahm, {\it {The construction of all Selfdual Multi-monopoles by the ADHM
  Method.}},  in {\em {Trieste Cent. Theor. Phys. - IC-82-016 (82,REC.MAR.) 8p,
  In *Trieste 1981, Proceedings, Monopoles In Quantum Field Theory*, 87-94 and
  Trieste Cent. Theor. Phys. - IC-82-016 (82,REC.MAR.) 8p}}, 1982.

\bibitem{Cheng:2013mla}
B.~Cheng and C.~Ford, {\it {Fermion Zero Modes for Abelian BPS Monopoles}},
  {\em Phys. Lett.} {\bf B720} (2013) 262--264,
  [\href{http://xxx.lanl.gov/abs/1302.2939}{{\tt arXiv:1302.2939}}].

\bibitem{VanBaal:2002rt}
P.~Van~Baal, {\it {Chiral zero mode for Abelian BPS dipoles}},  in {\em {NATO
  Advanced Research Workshop on Confinement, Topology, and other
  Nonperturbative Aspects of QCD Stara Lesna, Slovakia, January 21-27, 2002}},
  2002.
\newblock \href{http://xxx.lanl.gov/abs/hep-th/0202182}{{\tt hep-th/0202182}}.

\bibitem{Cecotti:1991me}
S.~Cecotti and C.~Vafa, {\it {Topological antitopological fusion}},  {\em Nucl.
  Phys.} {\bf B367} (1991) 359--461.

\bibitem{Cecotti:1992qh}
S.~Cecotti, P.~Fendley, K.~A. Intriligator, and C.~Vafa, {\it {A New
  supersymmetric index}},  {\em Nucl. Phys.} {\bf B386} (1992) 405--452,
  [\href{http://xxx.lanl.gov/abs/hep-th/9204102}{{\tt hep-th/9204102}}].

\bibitem{Cecotti:1992rm}
S.~Cecotti and C.~Vafa, {\it {On classification of N=2 supersymmetric
  theories}},  {\em Commun. Math. Phys.} {\bf 158} (1993) 569--644,
  [\href{http://xxx.lanl.gov/abs/hep-th/9211097}{{\tt hep-th/9211097}}].

\bibitem{Kronheimer:1985}
P.~Kronheimer, {\it {Monopoles and Taub-NUT metrics}},  {\em Transfer thesis,
  Oxford University} (1985).

\bibitem{Hitchin:1988hk}
N.~J. Hitchin and M.~K. Murray, {\it {Spectral Curves and the Adhm Method}},
  {\em Commun. Math. Phys.} {\bf 114} (1988) 463--474.

\bibitem{Nakajima:1990}
H.~Nakajima, {\it Monopoles and nahm's equations},  {\em Lectures Notes in Pure
  and Applied Mathematics} (1993) 193--193.

\bibitem{Diaconescu:1996rk}
D.-E. Diaconescu, {\it {D-branes, monopoles and Nahm equations}},  {\em Nucl.
  Phys.} {\bf B503} (1997) 220--238,
  [\href{http://xxx.lanl.gov/abs/hep-th/9608163}{{\tt hep-th/9608163}}].

\bibitem{Cherkis:1997aa}
S.~A. Cherkis and A.~Kapustin, {\it {Singular monopoles and supersymmetric
  gauge theories in three-dimensions}},  {\em Nucl. Phys.} {\bf B525} (1998)
  215--234, [\href{http://xxx.lanl.gov/abs/hep-th/9711145}{{\tt
  hep-th/9711145}}].

\bibitem{Hitchin:1982gh}
N.~J. Hitchin, {\it {Monopoles and Geodesics}},  {\em Commun. Math. Phys.} {\bf
  83} (1982) 579--602.

\bibitem{Hitchin:1983ay}
N.~J. Hitchin, {\it {On the Construction of Monopoles}},  {\em Commun. Math.
  Phys.} {\bf 89} (1983) 145--190.

\end{thebibliography}\endgroup

\end{document}